\def \yskip{\penalty-50\vskip3pt plus 3pt minus 2pt}
\def \reference{\par \yskip \noindent \hangindent .4in \hangafter 1}
\def \abc#1#2#3#4 {\reference#1, {\sl#2}, {\bf#3}, #4}
\def \blank {\lower 5pt\hbox to 0.75in{\hrulefill}}

\def \cm{~\rm{cm}}
\def \s{~\rm{s}}
\def \km{~\rm{km}}

\def \g{~\rm{g}}
\def \AU{~\rm{AU}}
\def \yrs{~\rm{yrs}}

\def \K{~\rm{K}}
\def \G{~\rm{G}}
\def \erg{~\rm{erg}}

\def \lae{\mathrel{<\kern-1.0em\lower0.9ex\hbox{$\sim$}}}
\def \gae{\mathrel{>\kern-1.0em\lower0.9ex\hbox{$\sim$}}}

\documentstyle[11pt,aasms,tighten,flushrt]{article}

\begin{document}
\small

\setcounter{page}{1}
\begin{center} \bf 
TURBULENT DYNAMO IN ASYMPTOTIC GIANT BRANCH STARS
\end{center}

\begin{center}
Noam Soker and Essam Zoabi$^1$\\
Department of Physics, University of Haifa at Oranim\\
Oranim, Tivon 36006, ISRAEL \\
soker@physics.technion.ac.il \\
$^1$ permanent address: St. Joseph High School, P.O. Box 99 
Nazareth 16100, Israel
\end{center}


\begin{center}
\bf ABSTRACT
\end{center}

 Using recent results on the operation of turbulent dynamos,
we show that a turbulent dynamo can amplify a large scale magnetic
field in the envelopes of asymptotic giant branch (AGB) stars.
 We propose that a slow rotation of the AGB envelope can
fix the symmetry axis, leading to the formation of an
axisymmetric magnetic field structure.
 Unlike solar-type $\alpha \omega$ dynamos, the rotation has only
a small role in amplifying the toroidal component of the magnetic
field; instead of an $\alpha \omega$ dynamo we have an
$\alpha^2 \omega$.
 The magnetic field can reach a value of
$B \simeq 10^{-4} B_e \simeq 0.01 \G$, where $B_e$ is
the equipartition (between the turbulent and magnetic energy densities)
magnetic field.
 The large-scale magnetic field is strong enough for the
formation of magnetic cool spots on the AGB stellar surface.
The spots can regulate dust formation, hence mass loss rate,
leading to axisymmetric mass loss and the formation of elliptical
planetary nebulae (PNe).
 Despite its role in forming cool spots, the large scale
magnetic field is too weak to play a dynamic role and directly
influence the wind from the AGB star, as required by some models.
 We find other problems in models where the magnetic field plays a
dynamic role in shaping the AGB winds, and argue that they cannot
explain the formation of nonspherical PNe.

\noindent
{\it Subject heading:}   
  Planetary nebulae: general
$-$ stars: AGB and post-AGB
$-$ stars: mass loss
$-$ stars: magnetic fields 
$-$ circumstellar matter


\section{INTRODUCTION}

 The axisymmetric structures, e.g., elliptical or bipolar, of most
planetary nebulae (PNe), has led many people to suggest that stellar
magnetic fields shape the winds from asymptotic giant branch (AGB)
star progenitors of PNe (e.g., Pascoli 1997;
Chevalier \& Luo 1994; Garcia-Segura 1997; Garcia-Segura {\it et al.}
1999; Matt {\it et al.} 2000; Blackman {\it et al.} 2001).
 Common to all these models and scenarios, hereafter termed
dynamic-magnetic models, is the dynamic role attributed to
the magnetic field.
 The different models are not identical in all their ingredients.
For example, in some models the shaping occurs close to the stellar
surface, through magnetic tension and/or pressure
(Pascoli 1997; Matt {\it et al.} 2000), while in others
(Chevalier \& Luo 1994; Garcia-Segura 1997)
the shaping occurs in the nebula at large distances from the star.
 Despite these and other differences the models seem to suffer from
the same basic problems, which, as we argue in the present paper,
prevent any of them from being the correct model for the shaping
of PNe.
Following earlier papers (Soker \& Harpaz 1992, 1999; Soker 1998) we
argue in section 2 of the present paper that magnetic fields are
not likely to play a dynamic role in shaping most PNe.
 Only in a minority of cases, where the progenitor AGB star was
substantially spun-up via a common envelope or tidal interaction,
it is possible, although not necessary, that the magnetic field had a
dynamic role in shaping the descendant PN.
 Our objections to these models, some raised in earlier papers
and some new, are summarized in section 2.

 One of the problems we find in some of the papers cited above is
that they scale properties from the solar model to AGB stars,
e.g., they assume that magnetic field can shape the intensive AGB
wind close to the stellar surface. 
 To emphasize the differences between the Sun and AGB stars regarding
magnetic activity and winds, we devote section 3 to a detailed comparison
of the relevant physical parameters between the Sun and AGB stars.
 These differences cannot be ignored in a consideration of
magnetic activity in AGB and post-AGB stars.

 In section 4 we show that magnetic activity can indeed take place in
AGB stars, but to a lesser degree than that required by the models
cited above.
 We argue that the magnetic field is most likely amplified
by a turbulent dynamo, an $\alpha^2 \omega$ dynamo where the main
role of the rotation is fixing a symmetry axis, rather than by a
solar type dynamo, the $\alpha \omega$ dynamo where the rotation
plays a crucial role in amplifying the toroidal component of
the magnetic field.
 Our estimate of the AGB magnetic activity is based on new papers
by Brandenburg and collaborators (Brandenburg 2001,
Brandenburg, Bigazzi, \& Subramanian 2001, and Brandenburg \& Dobler
2001, hereafter B2001, BBS, and BD01, respectively).
 As argued in earlier papers (e.g., Soker 1998; 2000) 
the magnetic field may become dynamically important at specific
locations near the surface where it forms cool spots, which can
regulate the mass loss process via the formation of dust above
these cool spots.
 However, the average magnetic energy density is much below the thermal
or kinetic energy density in the wind, hence the magnetic field has
no direct influence on the mass loss process.
 The much weaker magnetic activity than that required
in the dynamic-magnetic models is sufficient to explain
the following observations of magnetic fields in cool giant stars.  
 Kemball \& Diamond (1997) detected a magnetic field in the extended
atmosphere of the Mira variable TX Cam, with $B \lae 5 G$ 
at the locations of SiO maser emission at a radius of
$4.8 \AU \simeq 2 R_\ast$.
 Magnetic field of $\sim 1 {\rm mG}$ is detected in the
OH maser emission regions of U Herculis (Palen \& Fix 2000).
The detection of X-ray emission from a few M giants (H\"unsch {\it et al.} 
1998; see also Schr\"oder, H\"unsch, \& Schmitt 1998) also hints at
the presence of magnetic fields in giant stars, but weak.
 We summarize our main results in section 5.

\section{PROBLEMS WITH DYNAMIC-MAGNETIC MODELS}

 In the present section we list the fundamental problems we find in
models that attribute a dynamic role to the magnetic field.
\newline
\subsection {Too fast rotation and/or unrealistic angular
momentum distribution}
 In earlier papers (Soker \& Harpaz 1992; Soker 1998) it was shown that
dynamic-magnetic models must incorporate a binary companion to
spin-up the envelope, since single stars slow down
markedly on the AGB (Soker 2001). 
 The model of Pascoli (1997) was criticized by Soker (1998; 2001).
 The model proposed by Chevalier \& Luo (1994) and extended by
Garcia-Segura (1997; see also Garcia-Segura {\it et al.} 1999) was
shown by Soker (1998) to require a binary companion.
 Basically, this model is based on the tension of the toroidal
component of the magnetic field in the wind during the transition from
the AGB to the PN phase at large distances from the star.
 In contrast to other dynamic-magnetic models, in this one close to the
star the magnetic pressure and tension are negligible compared
with the ram pressure and thermal pressure of the wind. 
 Only when the wind hits the outer PN shell, which is the remnant of the
slow wind, and goes through a shock and slows down,
does the toroidal component becomes important and shape the nebula.

 The recent paper by Blackman {\it et al.} (2001) seems to suffer
the most from this problem, in that it assumes an unrealistic angular
momentum distribution.
 Blackman {\it et al.} (2001) propose that the magnetic field is
amplified via a solar type $\alpha \omega$ dynamo in the core-envelope
interface (an amplification of the magnetic field close to the AGB core
was already suggested by Pascoli 1997). 
 For an $\alpha \omega$ dynamo a relatively large angular velocity
gradient is required, which they assume is because each mass
shell conserves its angular momentum from the main sequence up to
the upper AGB.
 This assumption seems unrealistic, since a strong coupling
is expected in the convective AGB envelope.
Even in the radiative core the powerful weak-field MHD instability
(Balbus \& Hawley 1994) is likely to force a solid body rotation.
 The strong magnetic field obtained by Blackman {\it et al.} (2001)
will force a very fast solid body rotation, as we now show,
such that the uniformly rotating envelope will slow down very fast
with mass loss (Soker 2001).

 A curved magnetic flux tube embedded inside the envelope exerts
a tension-force per unit volume given by
\begin{equation}
f={{B^2}\over{4 \pi R_c}},
\end{equation}
where $B$ is the magnitude of the magnetic field and $R_c$ the radius
of curvature of the magnetic flux tube.
 At one point the flux tube rises to a height $h$, with an upward speed $v_u$,
during a time $t$ given by $t \sim h/v_u$
(we ignore the unlikely case where the flux tube is exactly circular
with the same radial positions around its entire circumference).
 The differential rotation results in an average (over the raising time $t$)
relative azimuthal velocity between the upper and lower parts of the tube
given by $v_y \simeq (r d \omega /dr) h/2$, where $d \omega /dr$ is the
angular velocity gradient.
 The flux tube will be azimuthally bent along a distance given by
$y \simeq v_y t \simeq 0.5 (r d \omega/dr) h t$, so that the radius
of curvature is \begin{equation}
R_c \sim {{h^2}\over{2 y}}
\simeq {{v_u}\over{r}} \left( {{d \omega}\over{dr}} \right)^{-1}. 
\end{equation}
 The azimuthal force resulting from the curved magnetic flux tube
exerts a moment per unit volume of $\sim r f$, where $f$ is given by
equation (1), and tends to bring the envelope to a uniform rotation.
 Equating this moment to the rate of change of angular momentum
per unit volume $\rho r^2 (d \omega/dt)$, and assuming that the average
magnetic flux tubes filling factor is $\beta$, gives the time required
to bring the envelope to a uniform rotation
\begin{equation}
\tau \equiv w \left( {{d \omega}\over{dt}} \right)^{-1}
\simeq
{{v_u r}\over {v_A^2}} 
\left( {{d \ln \omega}\over{d \ln r}} \right)^{-1} \beta^{-1},
\end{equation}
where $v_A= [B^2/(4 \pi \rho)]^{1/2}$ is the Alfven velocity.
 The rising speed of a flux tube is some fraction of the sound speed,
which Blackman {\it et al.} take to be the Alfven speed $v_u=v_A$.
Substituting typical values used by Blackman {\it et al.}
in their magnetic field amplification zone,  we find
\begin{equation}
\tau \simeq 5 \times 10^{-3} 
\left( {r}\over { R_\odot} \right)
\left( {v_A}\over {4 \km \s^{-1}} \right)^{-1}
\left( {d \ln \omega}\over{d \ln r} \right)^{-1} \beta^{-1} \yrs.
\end{equation}
 For the magnetic field to play a significant dynamic role,
as required by the model of Blackman {\it et al.}, the filling
factor $\beta$ cannot be too small, i.e., $\beta \gg 10^{-3}$,
and since they have ${d \ln \omega}/{d \ln r} \gtrsim 1$,
we find the slowing-down time, which is about the time required to
force a uniform rotation, to be $\tau \ll 10 \yrs$.
 As mentioned earlier, the powerful weak-field MHD instability which
operates for much weaker magnetic fields  (Balbus \& Hawley 1994)
is likely to force a solid body rotation as well.
 We therefore conclude that their assumed angular velocity profile
is unrealistic. 

 In this regard it is not clear why Blackman {\it et al.} assume that
the remnant WD can slow down by the interaction of its magnetic field
with the wind, while it will not slow down (according to their assumption)
by the interaction of the magnetic field with the much denser
envelope during the AGB phase.

\subsection {A Too Low Density Contrast and the Transition to Aspherical
Mass Loss} 
 In some papers there is no clear distinction between 
bipolar and elliptical PNe. 
 One of the reasons is that the equatorial to polar density 
contrast achieved in these models is very low.
 To achieve even a moderate density contrast the models
have to assume an unrealistically strong magnetic field
(e.g., Matt {\it et al.} 2000) or extremely fast rotation
(Garcia-Segura {\it et al.} 1999; see criticism by Soker \& Harpaz
1999). 
 In both these models, as well as that of Blackman {\it et al.}
(2001), there is no satisfactory explanation for the observations
that many PNe show a transition from an almost spherical to
highly non-spherical mass loss during the late stages of the
AGB and/or post-AGB. 
 Each model has to assume ad hoc that the relevant mechanism 
starts to operate only near the termination of the AGB, but no 
satisfactory physical mechanism is proposed for the switch-on
of the mechanism.

\subsection {No Radiative Dust-Acceleration}
 The common view, supported both by observation and theory,
is that the mechanism behind the intensive mass loss from
AGB stars is radiation pressure on dust
coupled with strong stellar pulsations
(e.g., Wood 1979; Jura 1986; Knapp 1986; 
Fleischer, Gauger \& Sedlmayr 1992;
Habing 1996;  Andersen, Loidl, \& H\"ofner 1999).
  However, the papers on dynamic-magnetic activity
(Matt {\it et al.} 2000; Blackman {\it et al.} 2001)
omit radiation pressure altogether. 
 For these models to work, the mass loss rate apparently has to 
be determined by direct magnetic activity, rather than by pulsation
and radiation pressure on dust.
 This is in contradiction with observations.
 
\subsection {A Too Strong X-Luminosity Is Predicted}

 In some of the dynamic-magnetic models the surface magnetic pressure
is comparable to the thermal pressure of the gas
(Matt {\it et al.} 2000; Blackman {\it et al.} 2001;
Garcia-Segura, Lopez, \& Franco 2001).
 This will lead to magnetic field reconnection, e.g., flares,
which will cause a very strong X-ray emission.
 For typical values of surface magnetic fields in these models,
$B \gtrsim 1 \G$, the expected X-ray luminosity is $\gtrsim 10^4$
times stronger than that of the Sun, if the reconnection rate per
unit surface area is similar to that in the Sun.
 If the X-ray luminosity is proportional to the optical luminosity
the same factor holds.
 Note that in the Sun, where the mass loss rate is determined by
magnetic activity, the average X-ray luminosity is of the same order
of magnitude as the rate of kinetic energy carried by the wind.
 The X-ray luminosity in the ROSAT/PSPC band is in the range of
$\sim 3 \times 10^{26}$ to $\sim 5 \times 10^{27} \erg \s^{-1}$,
at minimum and maximum, respectively (Peres {\it et al.} 2000).
 The solar wind's kinetic energy falls between these values.
 If this is the case for AGB stars in the dynamic-magnetic
models, the X-ray luminosity will be a factor of
$\sim 10^6-10^8$ stronger than in the Sun.
  This expectation, of $L_x \sim 10^{30} - 10^{35} \erg \s^{-1}$, in
dynamic-magnetic models is in sharp contradiction with observations.
 From observation, the maximum X-ray luminosities of red giant stars
are marginally larger than the solar X-ray luminosity
(Schr\"oder {\it et al.} 1998).
 In most cases $L_x< 10^{30} \erg \s^{-1}$, and further decreases in
late giants' evolution (H\"unsch \& Schr\"oder 1996).

 As mentioned in section 1, there are strong indication of
 magnetic fields around AGB stars.
 But we will argue in section 4 that these local fields 
(e.g., Palen \& Fix 2000 for U Herculis) result from a much
weaker magnetic activity.

 A final comment to the entire section is our view that while the
mechanism proposed by Chevalier \& Luo (1994; also Garcia-Segura 1997),
by Pascoli (1997), and by Garcia-Segura {\it et al.} (1999) may work,
but {\it only} if the progenitor AGB star is spun-up by a stellar
companion (although we still do not think these are the mechanisms for
shaping most PNe), we think that the scenarios proposed by
Matt {\it et al.} (2000) and Blackman {\it et al.} (2001) cannot work
at all.

\section{DIFFERENCES BETWEEN AGB STARS AND THE SUN}

  Soker (2000) discusses in detail a few major differences between dynamos
in main-sequence stars, e.g., the Sun, and any dynamo model
for AGB stars.
 To emphasis, clarify, and extend the list of these differences we present
them in Tab1e 1.
 The compared variables are listed in the first column, and
their symbols are given in the second column.
  The third column of Table 1 gives the units used, and the fourth and
fifth columns give the typical values in the Sun and in upper AGB stars,
respectively.
 The lower section of the Table gives variables relevant directly
to the magnetic activity.
 For AGB stars these variables have to be scaled with the angular
velocity $\omega$ and magnetic field intensity $B$.
 As representative values we take the orbital velocity to be
$0.001$ times the Keplerian angular velocity on the equator, i.e.,
$\omega=0.001 \Omega_3 \omega_{\rm kep}$,
and for the magnetic field we take $B=0.01 B_2 \G$.
 Angular velocity larger than the above scaling requires the
AGB star to be spun-up by a companion more massive than Jupiter
(to an order of magnitude; see Soker 2001 for the exact values).
 The magnetic field is scaled according to the results of the next
section.
 
The main relevant differences, and their implications, are as follows.
\newline
(1)  In the Sun the mass loss rate is determined mainly by the
magnetic activity.
 Therefore it is not surprising that the wind ram pressure near the
solar surface is smaller than the magnetic pressure (row number 18).
 In AGB stars, mainly during the final intensive wind (FIW; also called
superwind), the wind's ram pressure is larger by several orders of
magnitude than the magnetic field pressure, even if we take $B=1 \G$.
 This implies that magnetic field has no dynamic role in determining
the mass loss rate, and that mass leaving the star drags the magnetic
field lines, rather than being dragged by the magnetic field lines.

(2) In the Sun the dynamo number is $N_D > 1$ and the Rossby number
is ${\rm Ro} <1$ (to an order of magnitude $N_D \sim {\rm Ro}^{-2}$) 
as is required by standard $\alpha \omega$ dynamo models.
 The dynamo number is the square of the ratio of the magnetic field
amplification rate in the $\alpha \omega$ dynamo model, to the ohmic
decay rate. 
 In AGB stars, even if rotating close to the break-up velocity,
the opposite inequalities hold (rows 19 and 23).
 As noted by Soker (2000), the low value of the dynamo number in AGB
stars implies that the convective motion amplifies both the poloidal
and toroidal magnetic components, i.e., $\alpha^2$ dynamo.
 This is the subject of the next section. 

(3)  The fast convection motion in upper AGB stars
(row 9; see Soker \& Harpaz 1999 for details) means that the
energy density in the convective motion is comparable to the
thermal energy.
 In the Sun this ratio is very small.
 The strong convection also hints at the possible operation of a
turbulent dynamo in AGB stars. 

(4) As the AGB envelope gets depleted the mass loss time, defined as
$M_{\rm env}/\dot M$, where $M_{\rm env}$ is the envelope mass,
becomes shorter than the rotation period.
 In the Sun the mass loss time is longer by many orders of magnitude
than the rotation period. 
 The mass  loss time is not given in the table since it changes over
several orders of magnitude as the mass loss rate increases and
envelope mass decreases toward the termination of the AGB.

(5) Some structural differences between AGB envelopes and the
solar envelope are not presented here (see Soker \& Harpaz 1999;
note that the density scale given in their figs. 1-6 is off by a
factor of 10; the correct density scaling is in their figure 6).
 An example is the convective region, which in AGB stars
is very thick, whereas in the Sun its width is only $ 0.3 R_\odot$.

 The main conclusion from this section, which was already mentioned in
the previous section, is that the processes related to rotation,
convection, mass loss, and magnetic activity cannot simply be
scaled from the Sun to AGB stars.
 In particular, it is wrong to assume that the mass loss rate and/or
geometry are dynamically dictated by the magnetic activity.

\section{THE $\alpha^2\omega$ DYNAMO IN AGB STARS}
 The amplification of a large-scale magnetic field in AGB envelopes
via turbulent dynamo was mentioned before (Soker 2000), where simple
arguments in favor of an $\alpha^2 \omega$ dynamo were given.
 The main role of rotation in the $\alpha^2 \omega$ dynamo is fixing
a symmetry axis, rather than amplifying the azimuthal
component of the magnetic field.
 We use recent results by Brandenburg and collaborators
(B2001, BBS, and BD01), the most relevant for us being
as follows.
(1) The short mass loss time (see previous section) means that the
dynamo has open boundaries, with significant implications
for the amplification process (Blackman \& Field 2000; BBS).
(2) The turbulent dynamo ($\alpha^2$ dynamo) can amplify
a large-scale magnetic field (B2001, BD01).
(3) In a homogeneous and isotropic model, even a small perturbation
can lead to a preferred direction (B2001).
 In the present case the slow rotation of the AGB envelope
fixes the symmetry axis.
(4) With periodic boundary conditions the magnetic field energy reaches
equipartition with the turbulent energy, or even exceeds it (B2001).
(5) With open boundaries the magnetic energy density is lower
by a factor of $\sim \eta/\eta_{\rm eff}$ (BD01), where
$\eta$ is the magnetic diffusion coefficient (due to processes on
atomic scales), and $\eta_{\rm eff}$ is the effective magnetic diffusion 
coefficient due to convection.
 We now evaluate this ratio, and show that the expected magnetic
field is strong enough to form magnetic cool spots on the surface
of AGB stars.

 The derivation of the magnetic energy density follows the results
of BD01 (their $\S 5.1$) and B2001 (his $\S 3.6$).
 As stated above, with periodic boundary conditions the dynamo
brings the magnetic energy density to an equipartition with the turbulent
energy density, and even exceeds it by a factor of
$R_\ast/l_T$ (B2001 eq. 46), where $R_\ast$ is the size of the periodic
dynamo, taken here to be the stellar radius, and $l_T$ is the
forcing scale, taken here as the mixing length, i.e., about the
pressure scale height.
 We find that in AGB stars $R_\ast/l_T \simeq 5$, as in the calculations
of B2001. 
 Comparing eq. (23) of BD01 with equation (45) of B2001 gives the
magnetic energy density for a dynamo with open boundaries $E_{\alpha^2}$,
in terms of the equipartition magnetic energy density $E_e=E_T$, where
$E_T=\rho v_T^2$ is the turbulent kinetic energy density, 
\begin{equation}
E_{\alpha^2} = E_e {{\eta}\over{\epsilon_H \eta_{\rm eff}}}
{{R_\ast}\over{l_T}},
\end{equation}
where $\epsilon _H$ ($\epsilon_H \sim -3$ in the calculations of BD01)
is a nondimensional quantity defined in equation (17) of BD01.
 The effective magnetic diffusivity is given by equation (22) of BD01.
Since in AGB stars the turbulent magnetic diffusivity is much
larger than the magnetic diffusivity coefficient (see Table 1),
we can write to an order of magnitude
\begin{equation}
\eta_{\rm eff} \simeq \eta_T 
{{\epsilon_Q}\over{2 \pi \epsilon_H}} ,
\end{equation}
where $v_T=v_T l_T$, $v_T$ is turbulent velocity, and
$\epsilon_Q$ ($\vert \epsilon_Q \vert \sim 0.01-0.3$ in BD01)
is defined in equation (19) of BD01 as the ratio of
the magnetic helicity flux to the quantity $B^2 v_T$.
 The magnetic energy diffusivity coefficient due to processes on
atomic scales in AGB envelopes is
$\eta \sim 10^8 (T_{\rm env}/10^4 \K)^{-3/2} \cm^2 \s^{-1}$,
where $T_{\rm env}$ is the envelope temperature. 
 Substituting other typical values for upper AGB stars (e.g., 
Soker \& Harpaz 1999), $v_T\simeq 10 \km \s^{-1}$ and
$l_T \simeq 50 R_\odot$, we find
\begin{equation}
{{E_{\alpha^2}}\over{E_e}} \simeq  10^{-8}  
\left( {{\epsilon_Q}\over{0.1}} \right)^{-1}
\left( {{R_\ast}\over{10 l_T}} \right)
\left( {{T_{\rm env}}\over{10^4 \K}} \right) ^{-3/2}.
\end{equation}
 With the density in the outer convective region of AGB stars taken
to be $\rho=10^{-9} \g \cm^{-3}$, the equipartition magnetic pressure 
is $B_e=(8 \pi E_e)^{1/2} = (8 \pi \rho v_T^2)^{1/2} = 160 \G$.
 Taking the square root of equation (7) and using this value of
the equipartition magnetic field, we find the expected magnetic field 
intensity from an $\alpha^2$ dynamo in upper AGB stars to be 
\begin{equation}
{B_{\alpha^2}} \simeq  0.01  G.
\end{equation}
 This is two orders of magnitude lower than the required magnetic field
in most dynamic-magnetic models,  but it is of the same order of 
magnitude as the magnetic field required for the formation of  
magnetic cool spots in the magnetic cool spots model 
(Soker 1998, eq. 10; note that in that equation $\eta$ has a 
different meaning).

\section{SUMMARY}
 We analyzed recently proposed models which attribute the nonspherical
mass loss process from upper AGB stars to strong magnetic fields,
(dynamic-magnetic models)
and used recent results (B2001, BBS, BD01) on the operation
of a turbulent dynamo, i.e., an $\alpha^2$ dynamo, and obtained
the following results.
\newline
(1) We find ($\S 2$) problems in models where the magnetic field plays a
dynamic role in shaping the AGB winds close to the stellar surface,
i.e., the magnetic energy flux is of the same order of, or stronger
than, the wind's kinetic flux.
 From theoretical considerations we found that these models have to
assume that the AGB envelopes rotate at very high speeds, and/or to assume
unrealistic angular momentum distribution in the envelope.
 From observational considerations, the X-luminosity expected from
these models, due to magnetic field reconnection (as in the Sun),
is much higher than limits set by X-ray observations. 
\newline
(2) We argue that most likely the amplification of magnetic fields in
AGB stars is due to a turbulent ($\alpha^2$) dynamo, where the
azimuthal magnetic field component is amplified by convection, and not
by differential rotation as in the solar $\alpha \omega$ dynamo.
Only in a minority of AGB stars that have been substantially
spun-up by stellar companions (Soker 2000) can an $\alpha \omega$ model
be effective. 
 We found by applying the recent results of B2001 and BD01
to AGB stars that the expected average magnetic field is
$B_{\alpha^2} \simeq 10^{-4} B_e \simeq 0.01 \G$, where $B_e$ is
the equipartition magnetic field (where the magnetic energy density
is equal to the turbulent energy density).
 This field intensity is indeed much below the magnetic field
required by dynamic-magnetic models. 
\newline
(3) B2001 found that in a homogeneous and isotropic $\alpha^2$ dynamo
even a small perturbation can lead to a preferred direction.
 We proposed that a slow rotation of the AGB envelope can
fix the symmetry direction.
 For this role of the rotation, the dynamo in AGB stars is called
an $\alpha^2 \omega$ (Soker 2000).
\newline
(4) Although the large-scale magnetic field is much weaker than
the equipartition magnetic field, and cannot influence directly the
wind from the AGB star, it is strong enough for the formation of
magnetic cool spots on the AGB stellar surface (Soker 1998).
 As argued in earlier papers (e.g., Soker 1998; 2000) 
the magnetic field may become dynamically important in specific
locations near the surface where it forms cool spots.
 The magnetic cool spots can regulate the mass loss process via the
formation of dust above these cool spots, leading to axisymmetric
mass loss and the formation of elliptical PNe (Soker 1998).
                                                      
{\bf ACKNOWLEDGMENTS:} 
 This research was supported in part by grants from the
the US-Israel Binational Science Foundation.

\end{document}